ORIGINAL ARTICLE

# Histological insight into the hepatic tissue of the Nile monitor (*Varanus niloticus*)

**Yasser A. Ahmed**[1,*], **Mohammed Abdelsabour-Khalaf**[2], **Elsaysed Mohammed**[1]

[1]Department of Histology, Faculty of Veterinary Medicine, South Valley University, Qena, Egypt; [2]Department of Anatomy and Embryology, Faculty of Veterinary Medicine, South Valley University, Qena, Egypt

**ABSTRACT**

The liver of reptiles is considered an important study model for the interaction between environment and hepatic tissue. Little is known about the histology of the liver of reptiles. The aim of the current study was to elucidate the histological architecture of the liver of the Nile monitor *(Varanus niloticus)*. Liver fragments from the Nile monitor were collected in the summer season and processed for the light and electron microscopy. The liver of the Nile monitor was bi-lobed and the right lobe was found to be larger than the left lobe. Histological examination revealed indistinct lobulation of the liver, and the central vein, sinusoids and portal area were haphazardly organized. The hepatic parenchyma consisted of hepatocytes arranged in glandular-like alveoli or tubules separated by a network of twisted capillary sinusoids. The hepatocytes were polyhedral in shape with vacuolated cytoplasm and the nucleus was single rounded, eccentric, large and vesicular with a distinct nucleolus. The hepatocytes contained numerous lipid droplets, abundant glycogen granules and well-developed RER and mitochondria. The hepatocytes appeared to secrete into the bile canaliculi through the disintegration of their dark cytoplasm into the bile canaliculi. The space of Disse separating between the hepatocytes and sinusoids contained many recesses. The portal area contained branches of the portal vein, hepatic artery, bile duct and lymphatic vessels embedded in a connective tissue. Some non-parenchymal cells were described such as Kupffer cells, heterophils, melano-macrophages, intercalated cells, myofibroblasts in addition to the endothelium of the sinusoids. This is the first report about the histological structure of the liver of the Egyptian Nile monitor. The result presented here should be considered a baseline knowledge to compare with the pathological affections of the liver in this species.

## INTRODUCTION

The Nile monitor or *Varanus niloticus* is a large member of the monitor lizard family (*Varanidae*) found in many African countries. Its length ranges from one to two meters and weighs between 6-15 kg (Wikipedia 2017). It lives in close to Nile river, swamp and lakes in Egypt (Smith et al. 2008). The Nile monitor, like other reptiles, is an important animal for the ecosystem and acts as a scavenger for many non-vertebrate and small vertebrate animals (King and Green 1999). Its economic importance to human is increasing in many areas. For example, it is used as a pet animal, a source of food and leather to some African people and its high fat-content is a component of some medical treatments. Furthermore, it controls crocodile population through consuming of crocodile eggs. It also has a negative importance because it may attack livestock such as chickens in some areas (Szczepaniuk 2011).

Hibernation is a state of inactivity and decreasing the metabolic rate that associated with morphological and molecular changes of the tissue of the hibernator to adapt for special environmental conditions. The Nile monitor hibernates in large rock fishers or in deep holes in the mud close to the Nile river during the cold winter in Egypt (Wikipedia 2017).

Although many studies were applied on the geographical distribution and behavioral characteristics of the Nile monitor (Bennett 2002; Edroma and Ssali 1983; Enge et al. 2004; Mayes et al. 2005), only a few authors were interested in the morphological aspects of this animal. The structure of the alimentary tract of the Nile monitor is similar to other vertebrates (Jacobson 2007), with some specification (Ahmed et al. 2009).

Liver, the largest internal organ of the animal body, is reported to have a wide range of functions. It plays a vital role in the metabolism of absorbed nutrient from the intestine. The liver receives the intestinal nutrients through the portal vein, which conveys up to 80% of blood supply to the liver, while the rest of blood is supplied to the liver by the hepatic artery (Caceci 2015). The liver is essential for protein and glycogen synthesis and storage, bile synthesis and detoxification of metabolites (Young et al. 2014). In general, the liver parenchyma consists of hepatic lobules and portal areas. The hepatic lobules are made of hepatocytes arranged in regular plates surrounding a network of fenestrated capillaries known as sinusoids. Fenestrated sinusoids are lined with endothelial cells and Kupffer cells. The Kupffer cells are specialized phagocytic cells, which are needed for the removal of unwanted materials such as viral or bacterial infected cells and/ or apoptotic cells, in addition to their role in the liver repair (Dixon et al. 2013; Reuter 2007). Portal areas are embedded in the connective tissue between the hepatic lobules and contain branches of the portal vein, hepatic artery, bile duct and lymph vessels (Mescher and Junqueira 2016). Bile canaliculi are thin tubules collecting the bile secreted by the hepatocytes and run facing one side of hepatocytes. Bile canaliculi merge to form larger ductules and then main ducts at the portal areas emptying the bile into the gallbladder (Mescher and Junqueira 2016).

There is a gap in our understanding of the comparative liver histology between mammalian and non-mammalian vertebrates due to the lack of morphological studies of the liver from non-mammalian species, especially reptiles. One of the distinguishable histological features of the mammalian and non-mammalian liver is the arrangement of the hepatic parenchyma. For example, the hepatic parenchyma of mammalian liver contains one layer of hepatocyte arranged in plates or cords radiating from the central vein and lining the sinusoids (Reuter 2007). While in teleost, the hepatic plate is more than one-cell-layer thick around the sinusoids (Akiyoshi and Inoue 2004). Hepatocytes appeared in clusters and cords in birds or groups of single and clustered cells in reptiles (Odokuma and Omokaro 2015). Furthermore, in some reptiles, the hepatic tissue contains melano-macrophages; highly pigmented phagocytic cells, which is not found in the liver of mammalian or avian species (Firmiano et al. 2011; Henninger and Beresford 1990; Johnson et al. 1999). These melano-macrophages were described to have macrophage-like functions and may be used as monitors for environmental pollution (Agius and Roberts 2003). In some teleost, the exocrine pancreas is located within the liver and the organ is known as hepato-pancreas (Bertolucci et al. 2008; Petcoff et al. 2006).

The literature review showed the lack of reports on the normal histology of the liver of the Nile monitor. The current study aimed to describe the histological structure of liver of the Nile monitor with light and electron microscopy.

## MATERIALS AND METHODS

### Sample collection

Five apparently healthy adult male animals were used for the current study. The rostro-anal length and weight were $130 \pm 22$ cm $9 \pm 2$ kg, respectively. The animals were captured from a lake close to the Nile river in Qena city, Egypt, during the months of June to August 2017 (summer season). The animals were taken to the laboratory of the Histology Department in the Faculty of Veterinary Medicine, South Valley University (SVU), and left in a suitable controlled area for two days. On the third day, the animals were sacrificed, and the liver was excised. The gross morphology of the liver showed a normal picture with no pathological lesions. Fragments from the liver were rapidly fixed and processed for light and electron microscopic examination. The experiments have been approved by the ethics committee of the Faculty of Veterinary Medicine, SVU.

### Light microscopy

Small liver samples were rapidly fixed in 4% neutral buffered formalin, dehydrated in ascending grades of ethanol, cleared in xylol and embedded in paraffin wax blocks. Paraffin sections were cut to 5μm thickness using the fully automated Leica microtome (Leica RM2255, Germany) and stained with hematoxylin and eosin (H&E) as a general stain, Periodic Acid-Schiff (PAS) for detection of carbohydrates (glycogen), and Masson's trichrome to stain connective tissue according to the instructions (Jones et al. 2008). The sections were examined and photographed under a binocular microscope (Leica DMLS, Germany) coupled to a Leica digital camera (Leica ICC50, Germany).

### Electron microscopy

Liver samples were fixed in 2.5% glutaraldehyde, and then in 1% osmium tetroxide. For the transmission electron microscopy (TEM) samples were dehydrated in ascending grades of ethanol and embedded in Epon 812 resin (Electron Microscopy Sciences; USA). Resin-embedded samples were let to polymerize at 60Cº for 2-3 days. Semithin (0.5 μm) and ultrathin (80-100 nm) sections were taken with a glass knife. Semithin sections were stained with toluidine blue and examined with the light microscopy. Ultrathin sections were stained with lead citrate and uranyl acetate and examined with the TEM (JEOL1010, Japan). For the scanning electron microscopy (SEM), liver fragments were dehydrated in ascending grades of acetone, critical point dried, coated with aluminum-gold in a sputter coater and examined with the SEM (JEOL 5500, Japan). The TEM and the SEM examination were performed at the Central Laboratory of the SVU.

## RESULTS

### Light microscopy of the hepatic tissue of the Nile monitor (Fig. 1)

Grossly, the liver of the Nile monitor was a large dark brown bi-lobed organ and the right lobe was larger than the left one. On the dorsomedial surface, the lobes were partially separated by a narrow isthmus of hepatic tissue. A connective tissue capsule containing smooth muscle fibers surrounded the liver parenchyma (Fig.1A). The portal connective tissue area contained branches of the portal vein, artery and bile duct (Fig. 1A-C). The bile ducts were lined with simple cuboidal epithelium or simple columnar epithelium according to the degree of branching (Fig. 1C).

The central vein, sinusoids, and portal tissue were randomly distributed through the hepatic parenchyma (Fig. 1D). The hepatocytes grouped into glandular-like alveoli or tubules; each of them contained 3-8 cells and was surrounded by a network of twisted varied-size sinusoidal capillaries (Fig. 1C, D, E, H, I). With H&E, the hepatocytes were polyhedral in shape with vacuolated cytoplasm and single, large, rounded, eccentric and vesicular nucleus with prominent dark nucleolus (Fig. 1D, E, H, I). These vacuoles stained positive for lipids in semithin sections fixed in osmium tetroxide (Fig. 1F). The hepatocytes contained variable amounts of glycogen granules (Fig. 1G).

Many cells rather than hepatocytes could be observed with the light microscopy in the hepatic parenchyma. Phagocytic Kupffer cells were seen in the sinusoidal lumen or attached to their surface (Fig. 1D, F). Heterophils were found between the hepatocytes, inside the sinusoids or attached to the structural wall of the sinusoids. Heterophils characterized by needle-like acidophilic granulated-cytoplasm and a single, rounded, eccentric and basophilic nucleus (Fig. 1H). Melano-macrophages were occasionally seen in the hepatic parenchyma. They appear as clumps of cells with dark brown pigments (Fig.1I). The sinusoidal capillaries were lined with endothelial cells with a flattened nucleus (Fig. 1D). Many elliptical erythrocytes with oval nucleus appeared filling the sinusoids (Fig. 1D, E, H, I).

### Electron microscopy of the hepatic tissue of the Nile monitor (Fig. 2)

With the TEM, the hepatocytes had electron-dense cytoplasm and light nucleus rich in euchromatin with a distinct nucleolus and contained a variable number of different-sized lipid droplets, which appeared as non-membranous electron lucent rounded or oval structures (Fig. 2A). The cytoplasm contained well-developed RER (Fig. 2B, C) and numerous glycogen granules (Fig. 2C, D). Junctional desmosomes sealed off some parts of the lateral surfaces of the adjacent hepatocytes (Fig. 2C). The hepatocytes were surrounded by bile canaliculi at the lateral surfaces and many microvilli extended into the canalicular lumen (Fig. 2A-D). The hepatocytes appeared to secrete into the bile canaliculi by extrusion of their cytoplasmic contents into the bile canaliculi, and many parts of the cytoplasm appeared to cut off into the bile canaliculi (Fig. 2D). The Kupffer cells were pleomorphic and found inside the sinusoidal lumen or attached to the endothelial surface of the sinusoids (Fig. 2E). The Kupffer cells showed irregular large cellular body sending many processes or pseudopodia in the sinusoidal lumen and had large round and eccentric nucleus with many phagocytosed materials and apoptotic cells inside the cytoplasm (Fig. 2E). The sinusoid was lined by typical festered endothelial cells (Fig. 2F). The space of Disse locating between the hepatocytes and the sinusoid was filled with microvilli of the hepatocytes and amorphous materials (Fig. 2E, F). In some locations, the space of Disse extended into the intercellular spaces of the hepatocytes forming small recesses (Fig. 2E, G). We observed a lymphocyte-like cell with a large chromatin-patchy nucleus, sparse organelles and some cellular processes inside a recess extending from the space of Disse (Fig. 2G). A cell with a large dark nucleus and few cytoplasm, from which long processes extended underneath the endothelium of the sinusoids were seen in the space of the Disse (Fig. 2H).

With the SEM, the hepatocytes appeared as an interconnecting network of epithelial cells in the form of alveolar or tubular structures. The sinusoids appeared as a meshwork of tunnel spaces originating from the circular central vein and surrounded the hepatic tissue network (Fig. 2I). The lipid droplets were seen clearly bulging within the cytoplasmic mass of the hepatocytes (Fig. 2J). Branches of the portal vein, hepatic artery, and bile ducts were embedded in a connective tissue mass forming portal areas (Fig. 2I). The Kupffer cell processes were seen interrupted the sinusoidal wall and extending into the space of Disse (Fig. 2J). Numerous oval RBCs could be seen in the lumen of the sinusoids (Fig. 2J).

**DISCUSSION**

The current study was undertaken with the aim of exploring the histological features of the hepatic tissue of the Nile monitor. The histological examination revealed atypical organization of the hepatic tissue comparable to the higher vertebrates. The hepatic parenchyma did not show the known classical arrangement of the central vein, hepatic plates, sinusoids and portal area. The central vein, sinusoids and portal area were haphazardly organized. Furthermore, the hepatocytes appeared vacuolated and arranged in glandular-like alveoli or tubules surrounded by a twisted network of blood sinusoids. This hepatic morphology is not unique to the Nile monitor but was of a similar morphology to the liver of turtles (Mezyad 2015; Moura et al. 2009). The hepatocytes of the Nile monitor showed a high number of lipid droplets. Unlike, mammalian liver, the extensive lipid accumulation should not be considered a pathologic condition in such species. The high lipid content of the liver is likely related to the type of food consumption by this species, which includes invertebrates and small vertebrates (King and Green 1999). The specimens were collected from the liver during the summer (non-hibernating) season in Egypt. It is well known that the enzymes of lipogenesis in the liver of hibernating animals are higher in summer than winter (Anderson et al. 1989; Wang et al. 1997). Thus, the hepatocytes of the Nile monitor stores lipids in summer to be hydrolyzed during their hibernation in cool months of the winter as a source of energy.

The hepatocytes were arranged in alveoli or tubules, each of them contained 3-8 cells and separated by different-sized twisted sinusoids. The arrangement of hepatocytes is a species dependent. The hepatocytes in mammalian liver arranged in one-layered-thick plates (Treuting et al. 2017), in birds, the hepatocytes arranged into cords of 4-6 cells radiating from the central vein and in many fish, they arranged in two-layered-thick anastomosing cords separated by sinusoids (Bertolucci et al. 2008). The hepatocytes were surrounded by bile canaliculi at the lateral surfaces and many microvilli were seen in the bile canaliculi. Similarly, the bile canaliculi run parallel to the lateral surfaces in mammals (Baratta et al. 2009), but it was shown to be located at the apical surfaces of the hepatocytes in crocodiles (Van Wilpe 2013). The hepatocytes were seen secreting into the bile canaliculi through disintegration of their cytoplasm. Similar mode of secretion was described in the hypertrophic dark chondrocytes (Ahmed et al. 2007).

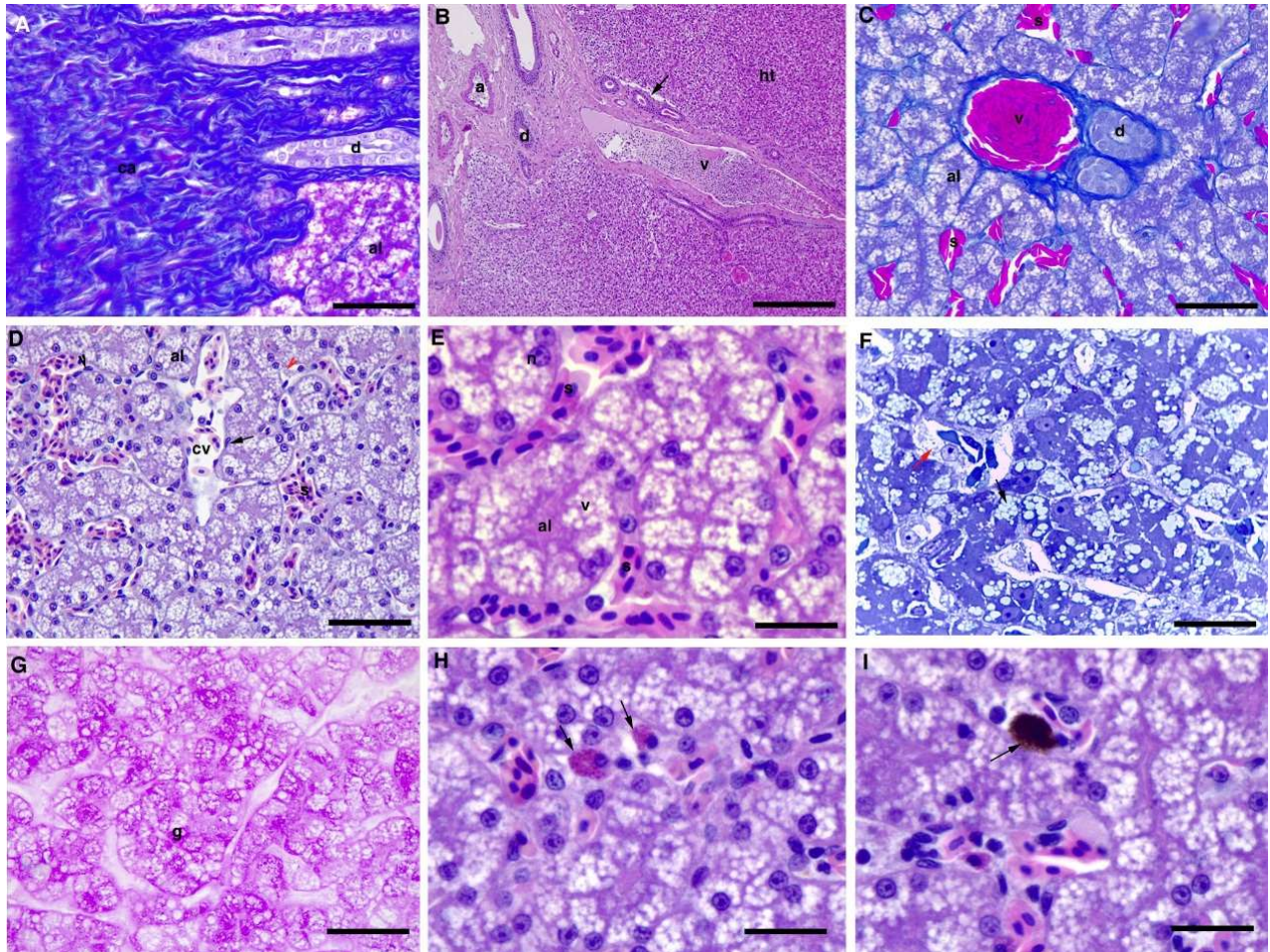

**Fig. 1: Light microscopy of the hepatic tissue of the Nile monitor.** Paraffin (A-E, G- I) and semithin (F) sections stained with Masson's trichrome (A, C), H&E (B, D, E, H, I), toluidine blue (F) and PAS (G). A: Liver capsule containing connective tissue (ca) and smooth muscle (red) fibers. Note; ducts (d) in the trabeculae and hepatic alveoli (al). B: Portal connective tissue containing branches of portal vein (v), hepatic artery (a), duct (d) and lymphatic vessels (arrow). Note; hepatic tissue (ht). C: Portal area containing ducts (d), vein (v), and hepatic alveoli (al) and sinusoids (s). D: Irregular distribution of the central vein (cv), sinusoids (s). Note; hepatic alveoli (al), endothelium of the central vein (black arrow), endothelium of the sinusoids (black arrowhead) and Kupffer cells (red arrowhead). E: Hepatic alveoli (al) consisting 3-8 hepatocytes. Note vacuolated cytoplasm (v), vesicular nucleolus with prominent nucleolus (n) of the hepatocytes and sinusoids (s). F: Lipid granules of the hepatocytes (black arrow). Note; Kupffer cells in the sinusoidal lumen (red arrow). G: Glycogen granules (g) inside the hepatocytes. H: Needle-like acidophilic granules of the heterophils (arrows). I: Pigments inside the melano-macrophages (arrow). Bars = 25 µm (A, C, D, F, G), 250 µm (B) and 62 µm (E, H, I).

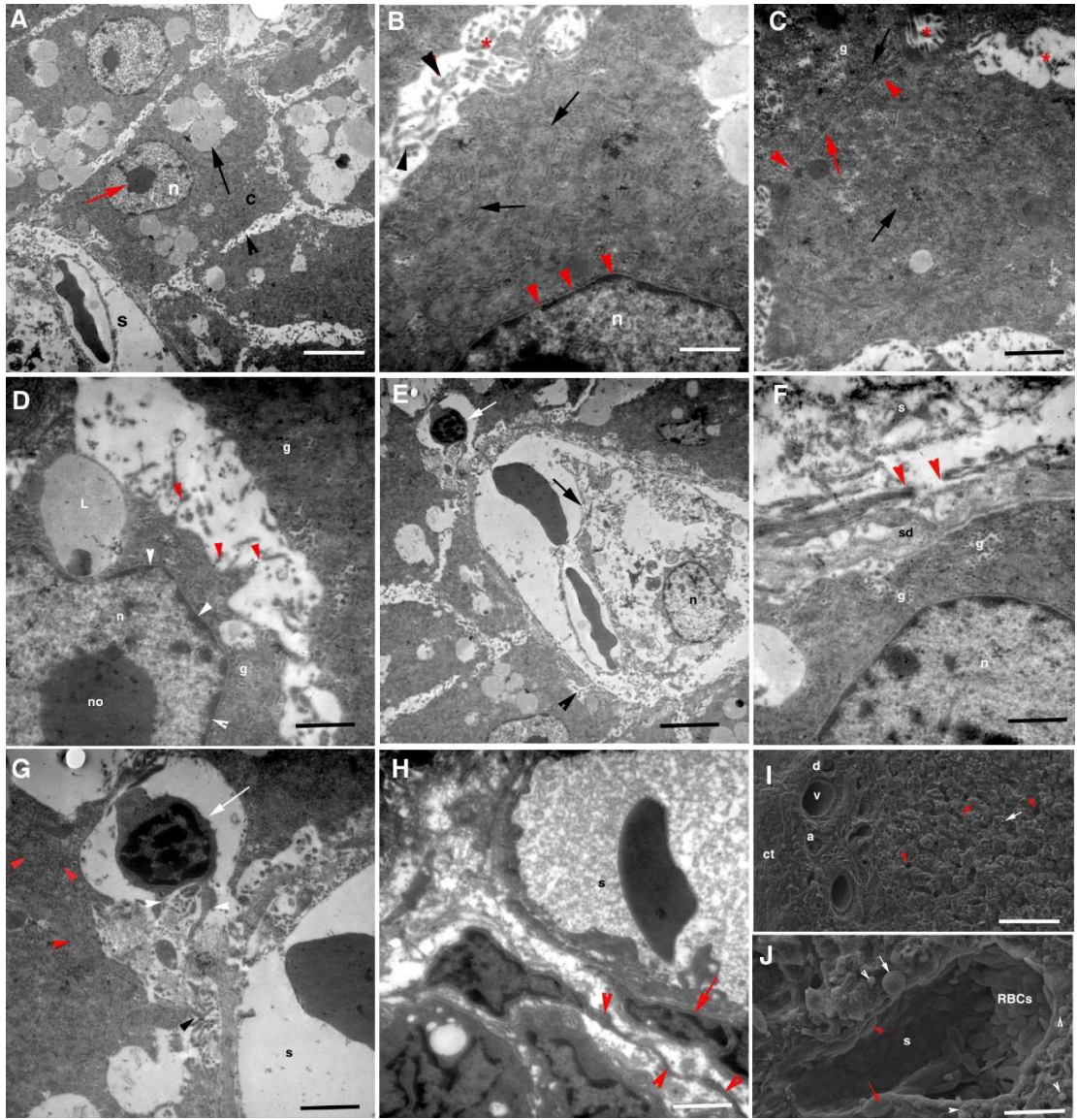

**Fig. 2: Electron microscopy of the hepatic tissue of the Nile monitor.** TEM (A-H) and SEM (I, J)-micrographs from the liver of the Nile monitor. A: Hepatocytes; note dark cytoplasm (c), light nucleus (n) with prominent nucleolus (red arrow), lipid droplets (black arrow), bile canaliculus (black arrowhead) and sinusoid (s). B: Hepatocytes containing well-developed RER (black arrows), membranes (red arrowheads) of the nucleus (n), microvilli (black arrowheads) protruded inside a bile canaliculus (red star). C: Parts of lateral surfaces of two hepatocytes (red arrow), sealed off with desmosomes (red arrowheads); note well-developed RER (black arrow), glycogen (g) and canaliculi (red stars). D: Extrusion of the cytoplasmic contents of the hepatocytes into the bile canaliculus (red arrowheads); note non-membranous electron-lucent lipid granule (L), glycogen granules (g), nucleus (n), nuclear membranes (white arrowheads) and dark nucleolus (no). E: Kupffer cell with many phagocytosed materials (black arrow); note nucleus (n) of the Kupffer cell, recess extended from the sub-sinusoidal space (black arrowhead) and intercalated cell inside a recess (white arrow). F: The space of Disse (sd) between the sinusoid and the hepatocytes. Note; interrupted endothelium (red arrowheads) of the sinusoid (s), and glycogen (g) and nucleus (n) of the hepatocytes. G: Intercalated cell (white arrow) with processes (white arrowheads). Not; microvilli of hepatocytes in the space of Disse (black arrowhead) and mitochondria of the hepatocytes (red arrowheads). H: Myofibroblast with long processes (red arrowheads); note endothelium (red arrow) of sinusoid (s). I: Portal area

consisting of connective tissue (ct) and containing branches of portal vein (v), hepatic artery (a) and bile duct (d); note hepatic alveoli (white arrow), central vein (red arrowheads) and sinusoid (red arrow). J: Sinusoid (s) filled with RBCs; note nucleus of the endothelium (red arrowhead), part of Kupffer cell (red arrow) extending from the sinusoid into the space of Disse, and nucleus (white arrow) and lipid droplets (white arrowheads) of hepatocytes. Bars = 400 nm (A), 166 nm (B), 200 nm (C) 166 nm (D), 400 nm (E), 166 nm (F), 200 (G), 200 nm (H), 25 μm (I) and 100 μm (J).

The non-parenchymal cells of the hepatic tissue of the Nile monitor included endothelium, Kupffer cells, heterophils, melano-macrophages and blood cells in addition to two cells with characteristic morphology; a cell with lymphocyte-like morphology and a cell with the long processes.

The endothelial cells of the sinusoids were flat cells with a flattened nucleus and of the fenestrated type. This is like other mammalian and non-mammalian species (Mezyad 2015; Moura et al. 2009; Odokuma and Omokaro 2015; Reuter 2007; Van Wilpe 2013).

The Kupffer cells are known to protect the hepatic tissue from invading harmful materials and toxins (Cakici and Akat 2013; Naito et al. 2004). In the current study, the Kupffer cells were found inside the lumen of the sinusoids and some cells penetrated the endothelium fenestrae into the space of Disse. In mammalian liver, the Kupffer cells along with the endothelium build up the wall of the fenestrated sinusoids and never extend inside the lumen (Naito et al. 2004). In Juvenile crocodile, the Kupffer cells were found in several locations; in the sinusoidal wall, inside the lumen or between the hepatic cells and the space of Disse (van Wilpe and Groenewald 2014).

Heterophils were seen in the hepatic tissue of the Nile monitor. These cells were reported as circulating blood cells in birds and reptiles, and were suggested to have a phagocytic activity (Claver and Quaglia 2009; Rowley and Ratcliffe 1988).

Melano-macrophages were occasionally seen among the hepatocytes. These cells considered as a type of hepatic defense system against cytotoxic substances and their increased number may reflect the rate of pollution (Fenoglio et al. 2005; Leknes 2007). Some authors consider the melano-macrophages to be a sub-type of the Kupffer cells (Van Wilpe 2013). The abundance of these cells is varied among non-mammalian vertebrates. They are numerous in amphibians (Moura et al. 2009) and form melano-macrophages centers in the hepatic tissue of teleost (Agius and Roberts 2003).

Characteristic nucleated oval RBCs of the Nile monitor were seen filling the sinusoids. In addition to its known role as oxygen-carrier; RBCs are recently considered as one of the non-immune cells supporting immune response against invading harmful substances (Passantino et al. 2007).

The space of Disse between the endothelium of the sinusoids and the hepatocytes extended into some recesses between the sinusoids and the hepatocytes. The recesses are likely to enhance the exposure of the hepatocytes to filtration by the blood circulation of the sinusoids. Furthermore, a cell with a patchy nuclear condensation and few cytoplasmic organelles was seen in a recess extending from the space of Disse. A cell with a similar morphology, termed "intercalated cell", was seen in the hepatic tissue of crocodiles and was suggested to have an immune function (Van Wilpe 2013). The authors suggested this cell to be a type of Kupffer cell moving freely from the sinusoidal lumen into the space of Disse. Furthermore, some authors described cells with a similar morphology to the intercalated cells and described them as "empty Ito cells" which have no lipid droplets (Taira and Mutoh 1981). Another cell with a long subendothelial process was seen in the space of the Disse. Similar cells were described as myofibroblast in the liver of crocodiles. These cells form a type of fibroblast and share some contractile functions due to their fibril contents (Van Wilpe 2013).

This is the first report on the light and electron microscopic structures of the Egyptian Nile monitor. The basic knowledge of the Nile monitor presented in the current study is so important for a better understanding and diagnosis of liver diseases of this species.

**CONFLICT OF INTERESTS**

The authors declare that they have no conflict of interests.

**ACKNOWLEDGMENTS**

The authors thank Ms. Fatma A. Khalil for assisting in sample collections and Prof. Antar Abdallah, Professor of English Education at Taibah University, Saudi Arabia, for language proofreading of the manuscript.

**Prof. Dr. Yasser A. Ahmed**
*Department of Histology*
*Faculty of Veterinary Medicine*
*South Valley University*
*Qena, Egypt*
*LiveDNA: 20.1825*

**Email:** yasser.ali@vet.svu.edu.eg
**Tel.:** +20965211223
**Fax:** +20965211223
**Mobile Number:** +201005455629